\documentclass[openright,a4paper,11pt]{scrartcl}
\usepackage{fancyhdr}

\usepackage[utf8]{inputenc}
\usepackage[ngerman,english]{babel}
\usepackage{listings}
\usepackage{url}

\usepackage{multicol}
\usepackage{datetime}
\newtimeformat{mytime}{\twodigit{\THEHOUR}:\twodigit{\THEMINUTE}}
\newdateformat{mydate}{\twodigit{\THEYEAR}.\twodigit{\THEMONTH}.\twodigit{\THEDAY}}

\usepackage{xcolor}
\usepackage{minibox}

\usepackage{todonotes}
\usepackage{amsmath}
\usepackage{tabularx}
\usepackage{amssymb}
\usepackage{multirow}
\usepackage{bussproofs}
\usepackage{booktabs}
\usepackage{bm}
\usepackage{stmaryrd}
\usepackage[inline]{enumitem}
\usepackage{tabularx}
\usepackage{mdframed}
\usepackage{minibox}
\usepackage{xparse}
\usepackage{tikz}

\usetikzlibrary{arrows,decorations.pathmorphing,backgrounds,fit,positioning,shapes.symbols,chains,shapes}
\usepackage{makecell}

\definecolor{darkgray}{rgb}{0.95,0.95,0.95}
\definecolor{darkgreen}{rgb}{0.05,0.70,0.0}
\definecolor{lstcap}{cmyk}{0.43,0.35,0.35,0.01}
\definecolor{lightgreen}{rgb}{0.55,0.70,0.55}
\definecolor{darkred}{rgb}{0.85,0.50,0.50}
\definecolor{dark-red}{rgb}{0.4,0.15,0.15}
\definecolor{dark-blue}{rgb}{0.15,0.15,0.4}
\definecolor{defblue}{rgb}{0.15,0.15,0.4}
\definecolor{medium-blue}{rgb}{0,0,0.5}

\usepackage{varwidth}

\usepackage[backend=bibtex,
babel=hyphen,
maxbibnames=10,
isbn=false,
doi=false,
backref=true,
mincrossrefs=100,
firstinits=true,
backref=false,
maxcitenames=1]{biblatex}

\usepackage{hyperref}
\hypersetup{    colorlinks, linkcolor={dark-red},
    citecolor={dark-blue}, urlcolor={medium-blue}
}

\usepackage[hyperref]{ntheorem}

\newcommand{\slist}[1]{\ensuremath{\overline{#1}}}
\newcommand{\denot}[1]{\ensuremath{\llbracket #1 \rrbracket}}
\newcommand{\impl}{\ensuremath{\Longrightarrow\,}}

\newcommand{\update}[2]{\ensuremath{[#1\mapsto{}#2]}}

\newcommand{\M}[1]{\ensuremath{\mathit{#1}}}
\newcommand{\pto}{\ensuremath{\rightharpoonup}}

\newcommand{\indu}[1]{\ensuremath{\textbf{\textit{#1}}}}

\newcommand{\dom}{\ensuremath{\textup{\textit{dom\,}}}}
\newcommand{\incl}{\ensuremath{\subseteq}}
\newcommand{\set}[1]{\ensuremath{\{#1\}}}

\newcommand{\injon}[2]{\ensuremath{#1\rightarrowtail{}#2}}

\newcommand{\nrule}[1]{\ensuremath{\textup{\textsc{#1}}}}
\newcommand{\nbrule}[1]{\ensuremath{\textup{\textbf{\textsc{#1}}}}}

\newcommand{\Exp}{\ensuremath{\M{Exp}}}
\newcommand{\Trm}{\ensuremath{\M{Term}}}
\newcommand{\Fun}{\ensuremath{\mathcal{F}}}
\newcommand{\Act}{\ensuremath{\mathcal{A}}}

\newcommand{\Var}{\ensuremath{\mathcal{V}}}
\newcommand{\Val}{\ensuremath{\mathbb{V}}}
\newcommand{\Valb}{\ensuremath{\mathbb{V}_{\bot}}}

\newcommand{\Evt}{\ensuremath{\mathcal{E}}}

\newcommand{\valtobool}{\ensuremath{\beta}}
\newcommand{\fv}{\ensuremath{\textup{fv}}}

\newcommand{\inlan}[1]{\ensuremath{\langle #1 \rangle}\,}

\newcommand{\ilIn}{\textsf{in}}
\newcommand{\ilReturn}[1]{#1}
\newcommand{\ilReturnAA}[2]{\inlan{#2}\ilReturn{#1}}
\newcommand{\ilGoto}[2]{#1\,#2}
\newcommand{\ilGotoAA}[3]{\inlan{#3}\ilGoto{#1}{#2}}
\newcommand{\ilLetRec}[4]{\textsf{fun}\,#1\,#2=#3\,\ilIn\,#4}
\newcommand{\ilLetRecA}[5]{\textsf{fun}\,#1\,#2\,:\,#5\,=#3\,\ilIn\,#4}
\newcommand{\ilLetRecAA}[6]{\inlan{#5}\ilLetRecA{#1}{#2}{#3}{#4}{#6}}

\newcommand{\ilLet}[3]{\textsf{let}\,#1=#2\,\ilIn\,#3}
\newcommand{\ilLetAA}[4]{\inlan{#4}\ilLet{#1}{#2}{#3}}
\newcommand{\ilIf}[3]{\textsf{if}\,{#1}\,\textsf{then}\,{#2}\,\textsf{else}\,{#3}}
\newcommand{\ilIfAA}[4]{\inlan{#4}\ilIf{#1}{#2}{#3}}
\newcommand{\ilEvent}[4]{\textsf{let}~#1=#2\,#3~\ilIn~#4}

\newcommand{\freenam}[1]{\ensuremath{\fv(#1)}}

\newcommand{\opeval}[2]{\ensuremath{\denot{#1}\,#2}}

\newcommand{\fstate}[3]{\ensuremath{#1~|~#2~|~#3}}
\newcommand{\fstatet}[3]{\ensuremath{#1&|\,#2&|\,#3}}
\newcommand{\fevals}{\ensuremath{\longrightarrow}}
\newcommand{\fevalsg}[2]{\stackrel{#2}{\fevals}_{#1}}

\newcommand{\fevalstg}[4]{\ensuremath{
\begin{array}{llll}
        & #1 \\
\stackrel{#4}{\fevals}_{#3} & #2
\end{array}
}
}
\newcommand{\fevalstgsl}[4]{\ensuremath{#1~~\stackrel{#4}{\fevals}_{#3}~~#2}}

\newcommand{\fevalst}[2]{\fevalstg{#1}{#2}{}{\tau}}
\newcommand{\ievalst}[2]{\fevalstg{#1}{#2}{I}{\tau}}

\newcommand{\trmg}[2]{\ensuremath{\Downarrow_{#2}{#1}}}

\newcommand{\Sim}{\ensuremath{\sim}}

\newcommand{\redsys}[5]{\ensuremath{(#1, #2, #3, #4, #5)}}
\newcommand{\LTS}[3]{\ensuremath{(#1, #2, #3)}}
\newcommand{\red}{\ensuremath{\fevals}}

\newcommand{\G}{\ensuremath{\gamma}}

\newcommand{\LT}{\ensuremath{\Lambda}}

\newcommand{\LC}{\ensuremath{\Lambda}}
\newcommand{\lv}{\ensuremath{X}}

\newcommand{\coh}[3]{\ensuremath{#1~\vdash\textbf{\textup{coh}}~{}#3}}
\newcommand{\cohCtx}[2]{\ensuremath{#2\vdash\textbf{\textup{coh}}\,{}#1}}

\newcommand\restr[2]{{  \left.\kern-\nulldelimiterspace   #1   \vphantom{\big|}   \right|_{#2}   }}
\newcommand{\restrict}[2]{\ensuremath{\lfloor#1\rfloor_{#2}}}

\newcommand{\agr}[3]{\ensuremath{#1,#2\models#3}}

\newcommand{\live}[3]{\ensuremath{#1\vdash\textbf{\textup{live}\,}{}#3:#2}}
\newcommand{\lives}[3]{\ensuremath{#1\vdash\textbf{\textup{live}}\,{}#3:#2}}
\newcommand{\livesA}[3]{\ensuremath{#1\vdash\textbf{\textup{live}}\,{}#3}}
\newcommand{\liveCtx}[2]{\ensuremath{#1\models#2}}

\newcommand{\ri}[3]{\ensuremath{#1\vdash\indu{inj}~#3}}

\tikzset{
node distance = 1cm, auto,font=\footnotesize,
every node/.style={node distance=5mm},
core/.style={rectangle, rounded corners, draw, inner sep=5pt, text width=20mm, minimum height=5mm, font=\footnotesize\sffamily},
cora/.style={rectangle, rounded corners, draw, inner sep=5pt, rectangle split, rectangle split parts = 2, text width=30mm, minimum height=5mm, font=\footnotesize\sffamily},
label/.style={rectangle, draw, inner xsep= 5pt, minimum height=0.6cm, font=\footnotesize\sffamily},
}
\tikzstyle{mybox} = [inner sep=5pt, inner ysep=5pt,rounded corners]

\newcommand{\ndef}[1]{\textbf{#1}}

\newcommand{\myref}[1]{\autoref{#1}}

\DeclareDataInheritance{proceedings}{inproceedings}{
\inherit[override=true]{booktitle}{booktitle}
}

\AtEveryBibitem{  \ifentrytype{inproceedings}{}{\clearlist{editors}}  \ifentrytype{inproceedings}{\clearlist{location}\clearfield{volume}\clearfield{series}}{}  \clearfield{booksubtitle}  \clearfield{day}  \clearfield{month}  \clearfield{endday}  \clearfield{endmonth}  \clearfield{endyear}}

\tikzset{
node distance = 1cm, auto,font=\footnotesize,
every node/.style={node distance=5mm},
core/.style={rectangle, rounded corners, draw, inner sep=5pt, text width=16mm, minimum height=5mm, font=\footnotesize\sffamily},
cora/.style={rectangle, rounded corners, draw, inner sep=5pt, rectangle split, rectangle split parts = 2, text width=15mm, minimum height=5mm, font=\footnotesize\sffamily},
label/.style={rectangle, draw, inner xsep= 5pt, minimum height=0.6cm, font=\footnotesize\sffamily},
}

\newcommand{\ann}[1]{#1}
\newcommand{\anni}[2]{#1\cdot#2}
\newcommand{\annii}[3]{#1\cdot#2,#3}

\newcommand{\alphai}[4]{\ensuremath{#1,#2\vdash#3 \,\sim_\alpha\, #4}}
\newcommand{\alphaie}[4]{\ensuremath{#1,#2\vdash_{\Exp}#3 \,\sim_\alpha\, #4}}

\newcommand{\rmp}{\ensuremath{\rho}}
\newcommand{\rmpp}{\ensuremath{d}}

  \let\llncssubparagraph\subparagraph
  \let\subparagraph\paragraph
   \let\subparagraph\llncssubparagraph

\bibliography{main}

\newsavebox{\topprooftreebox}
\newlength{\topprooftreewidth}

\fancypagestyle{fancyplain}{\fancyhf{} \fancyfoot[C]{\textsc{\thepage}} 

}
\renewcommand{\sectionmark}[1]{\markright{\thesection}}
\NewDocumentEnvironment{topprooftree}{m}  {\begin{lrbox}{\topprooftreebox}}  {\DisplayProof\end{lrbox}
\setlength{\topprooftreewidth}{\wd\topprooftreebox}    \begin{minipage}[b][][t]{\topprooftreewidth}      \nrule{\small #1}\\[1mm]\usebox{\topprooftreebox}
    \end{minipage}
  }

\NewDocumentEnvironment{topprooftreeb}{m}  {\begin{lrbox}{\topprooftreebox}}  {\DisplayProof\end{lrbox}
\setlength{\topprooftreewidth}{\wd\topprooftreebox}    \begin{minipage}[b][][b]{\topprooftreewidth}      \nrule{\small #1}\\[1mm]\usebox{\topprooftreebox}
    \end{minipage}
  }

\makeatletter
\newcommand{\shorteq}{  \settowidth{\@tempdima}{-}  \resizebox{\@tempdima}{\height}{=}}
\makeatother

\makeatletter
\newenvironment{btHighlight}[1][]
{\begingroup\tikzset{bt@Highlight@par/.style={#1}}\begin{lrbox}{\@tempboxa}}
{\end{lrbox}\bt@HL@box[bt@Highlight@par]{\@tempboxa}\endgroup}

\newcommand\btHL[1][]{  \begin{btHighlight}[#1]\bgroup\aftergroup\bt@HL@endenv}
\def\bt@HL@endenv{  \end{btHighlight}  \egroup
}
\newcommand{\bt@HL@box}[2][]{  \tikz[#1]{    \pgfpathrectangle{\pgfpoint{1pt}{0pt}}{\pgfpoint{\wd #2}{\ht #2}}    \pgfusepath{use as bounding box}    \node[anchor=base west, fill=orange!30,outer sep=0pt,inner xsep=1pt, inner ysep=0pt, rounded corners=3pt, minimum height=\ht\strutbox,#1]{\raisebox{0pt}{\strut}\strut\usebox{#2}};
  }}
\makeatother

\defbibenvironment{bibliography}
{\begin{enumerate}
{\setlength{\leftmargin}{\bibhang}\setlength{\itemindent}{-\leftmargin}\setlength{\itemsep}{\bibitemsep}\setlength{\parsep}{\bibparsep}}}
{\end{enumerate}}
{\item}

\definecolor{darkgray}{rgb}{0.95,0.95,0.95}
\theoremnumbering{arabic}
\theoremstyle{plain}
\theorembodyfont{\normalfont}

\newtheorem{theorem}{Theorem}

\theoremnumbering{arabic}
\theoremstyle{plain}
\theorembodyfont{\normalfont}

\newtheorem{lemma}{Lemma}

\theoremnumbering{arabic}
\theoremstyle{plain}
\theorembodyfont{\normalfont}

\newtheorem{definition}{Definition}

\theoremnumbering{arabic}
\theoremstyle{plain}
\theorembodyfont{\normalfont}

\theoremstyle{nonumberplain}
\theoremheaderfont{\itshape}
\theorembodyfont{\normalfont}
\theoremsymbol{\ensuremath{_\blacksquare}}
\qedsymbol{\ensuremath{_\blacksquare}}

\begin{document}

\renewcommand{\sectionautorefname}{Section}
\renewcommand{\subsectionautorefname}{Subsection}
\pagestyle{headings}  \title{A Linear First-Order Functional \\ Intermediate Language for Verified Compilers}
\author{Sigurd Schneider, Gert Smolka, Sebastian Hack\\[3mm]\small
Saarland University, Saarbrücken, Germany
}

\maketitle
\begin{abstract}
We present the linear first-order intermediate language IL for verified compilers.
IL is a functional language with calls to a nondeterministic environment.
We give IL terms a second, imperative semantic interpretation
and obtain a register transfer language.
For the imperative interpretation we establish a notion of live variables.
Based on live variables, we formulate a decidable property called coherence
ensuring that the functional and the imperative interpretation of a term coincide.

We formulate a register assignment algorithm for IL and prove its correctness.
The algorithm translates a functional IL program into an equivalent imperative IL program.
Correctness follows from the fact that the algorithm reaches a coherent program after consistently renaming local variables.
We prove that the maximal number of live variables in the initial program bounds the number of different variables in the final coherent program.
The entire development is formalized in Coq.
\end{abstract}

 \section{Introduction}
We study the intermediate language IL for verified compilers.
IL is a linear functional language with calls to a nondeterministic environment.

We are interested in translating IL to a register transfer language.
To this end, we give IL terms a second, imperative interpretation called IL/I.
IL/I interprets variable binding as assignment, and function application as \emph{goto},
where parameter passing becomes parallel assignment.

For some IL terms the functional interpretation coincides with the imperative interpretation.
We call such terms \emph{invariant}.
We develop an efficiently decidable property we call \emph{coherence} that is sufficient for invariance.
To translate IL to IL/I, translating to the coherent subset of IL suffices, i.e.
the entire translation can be done in the functional setting.

The notion of a live variable is central to the definition of coherence.
Liveness analysis is a standard technique in compiler construction to over-approximate the set of variables the evaluation of a program depends on.
Coherence is defined relative to the result of a liveness analysis.

\begin{figure}[h]
\begin{tabular}{p{\linewidth /2-2\tabcolsep}p{\linewidth /2-2\tabcolsep}}
\begin{lstlisting}[caption={},belowskip=-\baselineskip,aboveskip={-0.8\baselineskip}]
let i = 1 in
fun f (j,p) =
 let c = p <= m in
 if c then
  let k = p * j in
  let m = p + 1 in
  f (k,m)
 else
  j
in f (i,n)
\end{lstlisting}
&
\begin{lstlisting}[caption={},belowskip=-\baselineskip,aboveskip={-0.8\baselineskip}]
i := 1;
fun f (i,n) =
 c := n <= m;
 if c then
  i := n * i;
  n := n + 1;
  f (i,n)
 else
  i
in f (i,n)
\end{lstlisting}
\end{tabular}
\caption{Program (a) and (b) computing $ F(n,m) := n * (n+1) * \ldots * m $}
\label{fourprog}
\end{figure}

Inspired by the correspondence between SSA~\cite{DBLP:journals/toplas/CytronFRWZ91}
and functional programming~\cite{Kelsey:1995, DBLP:journals/sigplan/Appel98}, we formulate a register assignment algorithm~\cite{DBLP:conf/cc/HackGG06} for IL and show that it realizes the translation to IL/I.
For example, the algorithm translates program~(a) to program~(b).
Correctness follows from two facts:
First, register assignment consistently renames program~(a) such that the variable names correspond to program~(b).
Second, program~(b) is coherent, hence let binding and imperative assignment behave equivalently.
Parameter passing in IL/I can be eliminated by inserting parallel assignments~\cite{DBLP:conf/cc/HackGG06}.
In program~(b), all parameters $i,n$ can simply be removed, as they constitute self-assignments.

A key property of SSA-based register assignment is that the number of imperative registers required after register assignment is bounded by the maximal number of simultaneously live variables~\cite{DBLP:conf/cc/HackGG06}, which allows register assignment to be considered separate from spilling.
We show that our algorithm provides the same bound on the number of different variable names in the resulting IL/I term.

 \subsection{Related Work}
\label{chap:relwork}

Correspondences between imperative and functional languages were investigated already by \textcite{DBLP:journals/cacm/Landin65}.
The correspondence between SSA and functional programming is due to \textcite{DBLP:journals/sigplan/Appel98} and \textcite{Kelsey:1995} and consists of a translation from SSA programs to functional programs in continuation passing style (CPS)~\cite{DBLP:journals/lisp/Reynolds93,Appel92}.
\textcite{DBLP:journals/entcs/ChakravartyKZ03} reformulate SSA-based sparse conditional constant propagation on a functional language in administrative normal form (ANF)~\cite{DBLP:journals/lisp/SabryF93}.
Our intermediate language IL is in ANF, and a sub-language (up to system calls) of the ANF language presented in~\textcite{DBLP:journals/entcs/ChakravartyKZ03}.

Two major compiler verification projects using SSA exist.
{Comp\-Cert\-SSA} \cite{DBLP:conf/esop/BartheDP12} integrates SSA-based optimization passes into CompCert~\cite{Leroy-Compcert-CACM}.
VeLLVM \cite{DBLP:conf/popl/ZhaoNMZ12,DBLP:conf/pldi/ZhaoNMZ13}  is an ongoing effort to verify the production compiler LLVM~\cite{DBLP:conf/cgo/LattnerA04}.
Both projects use imperative languages with $\phi$-functions to enable SSA, and do not consider a functional intermediate language.
As of yet, neither of the projects verifies register assignment in the SSA setting.
In the non-SSA setting, a register allocation algorithm, which also deals with spilling, has been formally verified~\cite{DBLP:conf/esop/BlazyRA10}.

\textcite{DBLP:journals/entcs/BeringerMS03} use a language with a functional and imperative interpretation for proof carrying code.
They give a sufficient condition for the two semantics to coincide which they call Grail normal form (GNF).
GNF requires functions to be closure converted, i.e. all variables a function body depends on must be parameters.

\textcite{DBLP:conf/popl/Chlipala10} proves correctness for a compiler from Mini-ML to assembly including mutable references, but without system calls.
Register assignment uses an interference graph constructed from liveness information.
Chlipala restricts functions to take exactly one argument and requires the program to be closure converted prior to register assignment.
This means liveness coincides with free variables and values shared or passed between functions reside in an (argument) tuple in the heap: Effectively, register assignment is function local.
Chlipala does not prove bounds on the number of different variables used after register assignment and does not investigate the relationship to $\alpha$-equivalence.

 \subsection{Contributions and Outline}
\begin{itemize}
  \item We formally define the functional intermediate language IL and its imperative interpretation, IL/I.
    We establish the notion of live variables via an inductive definition.
    We identify terms for which both semantic interpretations coincide via the decidable notion of coherence.
  \item Inspired by SSA-based register allocation, we formulate a register assignment algorithm for IL and prove that it realizes an equivalence preserving transformation to IL/I. We show the size of the maximal live set bounds the number of names after register assignment.
    \item All results in this paper have formal Coq proofs, and the development is available online (see \myref{sec:formdev}).
We omit proofs in the paper for space reasons.
This version contains an appendix.
\end{itemize}
The paper is structured as follows:
We introduce the languages in \myref{sec:il_sem} and \myref{sec:ili}.
Program equivalence is defined in \myref{sec:progeq}.
We define invariance in \myref{sec:invariance}, establish a notion of live variables in \myref{sec:liveness}, and present coherence in \myref{sec:coherence}.
Register assignment is treated in \myref{sec:rassign}.

 \newcommand{\res}{\mathit{res}}
 \renewcommand\restr[2]{{  \left.\kern-\nulldelimiterspace   #1   \vphantom{\big|}   \right|_{#2}   }}

\newcommand{\extevt}{\ensuremath{\alpha}}
\newcommand{\evt}{\ensuremath{\phi}}
\newcommand{\configs}{\ensuremath{\Sigma}}
\newcommand{\config}{\ensuremath{\sigma}}
\newcommand{\closures}{\ensuremath{\mathcal{C}}}

\section{IL}
\label{sec:il_sem}
\paragraph{Values, Variables, and Expressions}
We assume a set $\Val$ of values and a function
$\valtobool:\Val\to\set{0,1}$ that we use to simplify the semantic rule for the conditional.
By convention, $v$ ranges over $\Val$.
We use the countably-infinite alphabet~$\Var$ for names $x,y,z$ of values, which we call \emph{variables}.

We assume a type $\Exp$ of expressions.
By convention, $e$~ranges over \Exp.
Expressions are pure, their evaluation is deterministic and may fail, hence
expression evaluation is a function $\opeval{\cdot}{}:\Exp\to(\Var\to\Valb)\pto\Valb$.
Environments are of type $\Var\to\Valb$ to track uninitialized variables.
We assume a function $\fv:\Exp\to\M{set}\,\Var$ such that for all environments $V,V'$
that agree on $\fv(e)$ we have $\opeval{e}{V}=\opeval{e}{V'}$.
We lift $\opeval{\cdot}{}$ pointwise to lists of expressions in a strict fashion: $\opeval{\slist{e}}$ yields a list of values if none of the expressions in~$\slist{e}$ failed, and $\bot$ otherwise.

\label{chap:il}
\paragraph{Syntax}
IL is a functional language with a tail-call restriction and system calls.
IL syntactically enforces a first-order discipline by using a separate
alphabet~$\Fun$ for names $f,g$ of function type, which we call \emph{labels}.
IL uses a third alphabet $\Act$ for names $\alpha$ which we call \emph{actions}.
The term $\ilLet{x}{\extevt}{\ldots}$ is like a system call $\alpha$ that non-deterministically returns a value. The formal development treats system calls with arguments.
Their treatment is straightforward and omitted here for the sake of simplicity.

IL allows function definitions, but does not allow mutually recursive definitions.
The syntax of IL is given in \myref{fig:ilf_syntax}.

\newcommand{\extexp}{\ensuremath{\eta}}
\begin{figure}[t]
  \centering
\begin{align*}
\eta ::=&~e~|~\extevt&&\quad\textup{extended expression}\\
\Trm\ni{}s,t ::=&~\ilLet{x}{\extexp}{s}&&\quad\textup{variable binding}\\
   |&~\ilIf{e}{s}{t}&&\quad\textup{conditional}\\
   |&~e&&\quad\textup{value}\\
   |&~\ilLetRec{f}{\slist{x}}{s}{t}&&\quad\textup{function definition}\\
   |&~f\,\slist{e}&&\quad\textup{application}
\end{align*}
  \caption{Syntax of IL}
  \label{fig:ilf_syntax}
\end{figure}

\newcommand{\bnfeq}{\ensuremath{\mathrel{:\mkern-2mu:\!\!\shorteq}}}
\newcommand{\defeq}{\ensuremath{\mathrel{:\!\!\shorteq}}}

\begin{figure*}[ht]

\begin{center}
\begin{topprooftree}{Op}
   \AxiomC{$\denot{e}\,V=v$}
      \UnaryInfC{\fevalst{\fstatet{F}{V}{\ilLet{x}{e}{s}}}
                      {\fstatet{F}{V\update{x}{v}}{s}}
                      }
\end{topprooftree}
\begin{topprooftree}{Cond}
   \AxiomC{$\denot{e}\,V=v$}
   \AxiomC{$\mathbb{\beta}(v) = i$}
   \LeftLabel{{\small\nrule{}}}
   \BinaryInfC{\fevalst{\fstatet{F}{V}{\ilIf{e}{s_0}{s_1}}}
                       {\fstatet{F}{V}{s_i}}}
\end{topprooftree}

\end{center}

\begin{center}
\begin{topprooftree}{Extern}
   \AxiomC{$v\in\Val$}
   \UnaryInfC{\fevalstgsl{\fstate{F}{V}{\ilEvent{x}{\alpha}{\!}{s}}}
                        {\fstate{F}{V\update{x}{v}}{s}}
                        {}{v=\alpha}}
\end{topprooftree}
\end{center}

\begin{center}
\begin{topprooftree}{Let}
  \AxiomC{}
  \UnaryInfC{\fevalst{\fstatet{F}{V}{\ilLetRec{f}{\slist{x}}{s}{t}}}
                     {\fstatet{F;f:(V, \slist{x}, s)}{V}{t}}}
\end{topprooftree}
\begin{topprooftree}{App}
  \AxiomC{$\denot{\slist{e}}\,V=\slist{v}$}
  \AxiomC{$F f = (V', \slist{x}, s)$}
  \BinaryInfC
    {\fevalst{\fstatet{F}{V}{\ilGoto{f}{\slist{e}}}}
             {\fstatet{F^f}{{V'}\update{\slist{x}}{\slist{v}}}{s}}}
\end{topprooftree}
\end{center}
\caption{Semantics of IL}
\label{fig:ili-sem}
\end{figure*}

\paragraph{Semantics}

\label{sec:ilf}
The semantics of IL is given as small-step relation $\fevals$ in \myref{fig:ili-sem}.
Note that the tail-call restriction ensures that no call stack is required.
The reduction relation $\fevals$ operates on \ndef{configurations} of the form $(F,V,s)$ where $s$ is the IL term to be evaluated.
The semantics does not rely on substitution, but uses an environment $V:\Var\to\Valb$ for variable definitions and a context $F$ for function definitions.
Transitions in~$\fevals$ are labeled with \emph{events} $\evt$.
By convention, $\psi$ ranges over events different from~$\tau$.
 $$\Evt\ni\evt ::= \tau~|~v=\alpha$$

A \ndef{context} is a list of named definitions.
A definition in a context may refer to previous definitions and itself.
Notationally, we use contexts like functions:
If a context $F$ can be decomposed as $F_1;f:a;F_2$ where $f\not\in\dom{F_2}$, we write $F f$ for $a$ and $F^f$ for $F_1;f:a$. Otherwise, $F f=\bot$.
To ease presentation of partial functions, we treat $f:\bot$ as if $f$ was not defined, i.e. $f\not\in\dom{(f:\bot)}$.
We write $\emptyset$ for the empty context.

A \ndef{closure} is a tuple $(V,\slist{x},s)\in\closures$ consisting of an environment $V$, a parameter list $\slist{x}$, and a function body $s$.
Since a function~$f$ in a context $F;f:\ldots;F'$ can refer to function definitions in $F$ (and to itself), the first-order restriction allows the closures to be non-recursive: function closures do not need to close under labels.
An application $f \slist{e}$ causes the function context $F$ to rewind to $F^f$, i.e. up to the definition of $f$ (rule \nrule{App}).
In contrast to higher-order formulations, we do not define closures mutually recursively with the values of the language.

A \textbf{system call} $\ilEvent{x}{\alpha}{}{s}$ invokes a function~$\alpha$ of the system, which is not assumed to be deterministic.
This reflects in the rule \nrule{Extern}, which does not restrict the result value of the system call other than requiring that it is a value.
The semantic transition records the system call name $\alpha$ and the result value $v$ in the event $v=\alpha$.

IL is \textbf{linear} in the sense that the execution of each term either passes control to a strict subterm,
or applies a function that never returns.
This ensures no run-time stack is required to manage continuations.
\textsc{While}, by contrast, uses sequentialization \texttt{;} to manage a stack of continuations.

\section{Imperative Interpretation of IL: IL/I}
\label{sec:ili}
We are interested in a translation of IL to an imperative language that does not require function closures at run-time.
We introduce a second semantic interpretation for IL which we call IL/I to investigate this translation.
IL/I is an imperative language, where variable binding is interpreted as imperative assignment.
Function application becomes a \emph{goto}, and parameter passing is a parallel assignment to the parameter names.
Closures are replaced by blocks $(\slist{x},s)\in\mathcal{B}$ and blocks do not contain variable environments.
Consequently, a called function can see all previous updates to variables.
For example, the following two programs each return 5 in IL/I, but evaluate to 7 in IL:

\noindent
\begin{minipage}{.5\textwidth}
\begin{lstlisting}[caption={}]
let x = 7 in
fun f () = x in
let x = 5 in f ()
\end{lstlisting}
\end{minipage}
\begin{minipage}{.45\textwidth}
\begin{lstlisting}[caption={}]
let x = 7 in
fun f () = x in
fun g x = f() in
let y = 5 in g y
\end{lstlisting}
\end{minipage}

To obtain the IL/I small-step relation $\fevals_{I}$, we replace the rules \nrule{F-Let} and \nrule{F-App} by the following rules:

\begin{center}
\begin{topprooftree}{I-Let}
  \AxiomC{$$}
  \UnaryInfC{$\ievalst{\fstatet{L}{V}{\ilLetRec{f}{\slist{x}}{s}{t}}}
    {\fstatet{L;f:(\slist{x}, s)}{V}{t}}$}
\end{topprooftree}
\begin{topprooftree}{I-App}
  \AxiomC{$\denot{\slist{e}}\,V=\slist{v}$}
  \AxiomC{$L f = (\slist{x}, s)$}
  \BinaryInfC{$\ievalst{\fstatet{L}{V}{\ilGoto{f}{\slist{e}}}}
                      {\fstatet{L^f}{{V}\update{\slist{x}}{\slist{v}}}{s}}$}
\end{topprooftree}
\end{center}

 \section{Program Equivalence}
\label{chap:progeq}
\label{sec:prog_eq}
\label{sec:progeq}
To relate programs from different languages, we abstract from
a configuration's internal behavior and only consider
interactions with the environment (via system calls) and termination behavior.
IL's reduction relation forms a labeled transition
 system (LTS) over configurations.

\begin{definition}
A \emph{reduction system} (RS) is a tuple
$\redsys{\configs}{\Evt}{\red}{\tau}{\res}$, s.t.
\begin{multicols}{2}
\begin{enumerate}[label=\textup{(\arabic*)}]
\item $\LTS{\configs}{\Evt}{\red}$ is a LTS
\item $\tau\in\Evt$
\item $\res:\configs\to\Valb$
\item $\res\,\config=v \Rightarrow \config~\textup{$\red$-terminal}$
\end{enumerate}
\end{multicols}
\noindent
An \emph{internally deterministic} reduction system (IDRS) additionally satisfies
\begin{enumerate}[label=\textup{(\arabic*)}]
\setcounter{enumi}{4}
\item $\config\fevalsg{}{\phi}\config_1 \land \config\fevalsg{}{\phi}\config_2 \Rightarrow \config_1 = \config_2$ \hfill \textup{action-deterministic} \quad

\item $\config\fevalsg{}{\phi}\config_1 \land \config\fevalsg{}{\tau}\config_2 \Rightarrow \phi = \tau$ \hfill \textup{$\tau$-deterministic} \quad
\end{enumerate}

\end{definition}

\newcommand{\ParTrace}{\ensuremath{\Pi}}
\subsection{Partial Traces}
\newcommand{\prods}{\ensuremath{\triangleright}}
\newcommand{\nil}{\ensuremath{\epsilon}}
We consider two configurations in an IDRS equivalent, if they produce the
same partial traces.
A partial trace $\pi$ adheres to the following grammar:
\begin{align*}
  \Pi\ni\pi ::= \nil~|~v~|~\bot~|~\psi\pi\end{align*}
We inductively define the relation $\prods\incl\configs\times\ParTrace$ such that
$\config\prods \pi$ whenever $\config$ produces the trace $\pi$.
In the following, we write trace for partial trace.

\begin{center}\begin{topprooftree}{Tr-Tau}
  \AxiomC{$\config \fevalsg{}{\tau} \config'$}
  \AxiomC{$\config' \prods \pi$}
  \BinaryInfC{\raisebox{0pt}[10pt]{$\config \prods \pi$}}
\end{topprooftree}\begin{topprooftree}{Tr-End}
  \AxiomC{\phantom{$\config \not\fevals$}}
  \UnaryInfC{\raisebox{0pt}[10pt]{$\config \prods \nil$}}
\end{topprooftree}
\begin{topprooftree}{Tr-Trm}
  \AxiomC{$\config~\textup{$\red$-terminal}$}
  \UnaryInfC{\raisebox{0pt}[10pt]{$\config \prods \res\,\config$}}
\end{topprooftree}
\begin{topprooftree}{Tr-Evt}
  \AxiomC{$\config \fevalsg{}{\psi} \config'$}
  \AxiomC{$\config' \prods \pi$}
  \BinaryInfC{$\config \prods \psi,\pi$}
\end{topprooftree}
\end{center}
\newcommand{\Prefixes}{\ensuremath{\mathcal{P}}}
The traces a configuration produces are given as
 $ \Prefixes{\config}=\set{\pi~|~\config\prods \pi}$.

\newcommand{\PrefixEq}{\ensuremath{\simeq}}
\begin{definition}[Trace Equivalence]
$\config \PrefixEq \config' :\!\iff \Prefixes{\config}=\Prefixes{\config'}$
\end{definition}

\begin{lemma}
 $\config$ silently diverges if and only if $\Prefixes{\config}=\set{\nil}$.
\end{lemma}

\subsection{Bisimilarity}
\label{sec:bisimilarity}
We give a sound and complete characterization of trace equivalence via bisimilarity.
Bisimilarity enables coinduction as proof method for program equivalence, which is more concise than arguing about traces directly.
We say a configuration $\config$ is \emph{ready} if the next step is a system call.
We write
$\config_2 \stackrel{R}{\leadsto}\config_1$ for
$\forall \config_1', \config_1\stackrel{\phi}{\fevals}\config_1' \Rightarrow \exists \config'_2, \config_2\stackrel{\phi}{\fevals}\config_2' \land \config_1'\mathrel{R}\config_2'$.
We write $\config\trmg{w}{}$ (where $w\in\Valb$) if $\config\red^\ast\config'$ such that $\config'$ is $\red$-terminal and $\res(\config')=w$.

\begin{definition}[Bisimilarity] Let $\redsys{S}{\Evt}{\fevalsg{}{}}{\res}{\tau}{}$ be an IDRS.
Bisimilarity $\Sim~\incl{S}\times{S}$ is coinductively defined as the greatest relation closed under the following rules:
\label{def:bisim}
\begin{center}
\begin{topprooftree}{Bisim-Silent}
  \AxiomC{$\config_1\fevalsg{}{}^{+}\config_1'$}
  \AxiomC{$\config_2\fevalsg{}{}^{+}\config_2'$}
  \AxiomC{$\config_1'\Sim\config_2'$}
  \doubleLine
  \TrinaryInfC{$\config_1\Sim\config_2$}
\end{topprooftree}
\begin{topprooftree}{Bisim-Term}
  \AxiomC{$\config_1\trmg{w}{}$}
  \AxiomC{$\config_2\trmg{w}{}$}
  \doubleLine
  \BinaryInfC{$\config_1\Sim\config_2$}
\end{topprooftree}
\end{center}
\begin{center}
\begin{topprooftree}{Bisim-Extern}
  \Axiom$\fCenter\config_1\fevalsg{}{}^{*}\config_1'$
    \Axiom$\fCenter\config_2\fevalsg{}{}^{*}\config_2'$
  \Axiom$\fCenter\config_1', \config_2' ~ \textup{ready}$
  \Axiom$\fCenter\config'_1 \stackrel{\Sim}{\leadsto}\config'_2$
  \Axiom$\fCenter\config'_2 \stackrel{\Sim}{\leadsto} \config'_1$
  \doubleLine
  \QuinaryInfC{$\config_1\Sim\config_2$}
\end{topprooftree}
\end{center}
\end{definition}
\nrule{Bisim-Silent} allows to match finitely many steps on both sides, as long as all transitions are silent.
This makes sense for IDRS, but would not yield a meaningful definition otherwise.
\nrule{Bisim-Extern} ensures that every external transition of $\config_1'$ is matched by the same external transition of $\config_2'$, and vice versa.
This ensures that if two programs are in relation, they react to every possible result value of the external call in a bisimilar way.
The premises that $\config_1',\config_2'$ are ready is there to simplify case distinctions by ensuring that the next event cannot be $\tau$.

\begin{theorem}[Soundness and Completeness]
Let $\redsys{S}{\Evt}{\fevalsg{}{}}{\res}{\tau}$ be an IDRS and $\config, \config' \in S$. Then:
$ \config \Sim \config' ~\iff~ {\config}\PrefixEq{\config'} $
\end{theorem}

The semantics of IL and of IL/I each forms an IDRS. We define $\res$ such that $\res(\config)=v$ if $\config$ is of the form $(F,V,e)$ and $\denot{e}\,V=v$. Otherwise, $\res(\config)=\bot$. The definitions for IL/I are analogous.
To relate configurations IL to IL/I, we form a reduction system on the sum $\configs_F + \configs_I$ of the configurations and lift $\red$ and $\res$ accordingly.
It is easy to see that the resulting reduction system is internally deterministic.
If not clear from context, we use an index $\config_F$, $\config_I$ to indicate which language a configuration belongs to.

 \section{Invariance}
\label{sec:invariance}
We call a term \emph{invariant} if it has the same traces in both the functional and the imperative interpretation.

\begin{definition}[Invariance]
\label{def:invariance}
A closed program $s$ is invariant if$$\forall\, V,~(\emptyset,V,s)_F \PrefixEq (\emptyset,V,s)_I$$\end{definition}
Invariance is undecidable.
We develop a syntactic, efficiently decidable criterion sufficient for invariance, which we call coherence.
Coherence simplifies the translation between IL and IL/I.

Coherence is based on the observation that some IL programs do not really depend on information from the closure.
Assume $F f = (V', \slist{x}, s)$ and consider the following IL reduction according to rule \nrule{App}:
$$    (F, V, \ilGoto{f}{\slist{e}})~~\fevalsg{}{}~~
             (F^f,V'\update{\slist{x}}{\slist{v}}, s)$$
If $V$ agrees with~$V'$ on all variables $X$ that $s$ depends on, then the configuration could have equivalently reduced to $(F^f,V\update{\slist{x}}{\slist{v}},s)$.
This reduction does not require the closure $V'$ and is similar in spirit to the rule \nrule{I-App}.
Coherence is a syntactic criterion that ensures $V$ and $V'$ agree on a suitable set $X$ at every function application.
We proceed in two steps:
\begin{enumerate}
\item \myref{sec:liveness} introduces the notion of live variables, which identifies a set that contains all variables a program depends on.
\item \myref{sec:coherence} gives the inductive definition of coherence and shows that coherent programs are invariant.
\end{enumerate}

 \section{Liveness}
\label{chap:liveness}
\label{sec:liveness}
A variable $x$ is \emph{significant} to a program~$s$ and a context~$L$,
if there is an environment~$V$ and a value~$v$ such that $(L,V,s)_I\not\PrefixEq (L,V[x\mapsto v],s)_I$.
Significance is not decidable, as it is a non-trivial semantic property.

Liveness analysis is a standard technique in compiler construction to over-approximate the set of variables significant to the evaluation of an imperative program.
While usual characterizations of live variables rely on data-flow equations~\cite{Nipkow:2014},
we define liveness inductively on the structure of IL's syntax.
To the best of our knowledge, such an inductive definition is not in literature.
The inductive definition factorizes the correctness aspect from the algorithmic aspect of liveness analysis.

We embed liveness information in the syntax of IL by introducing annotations
for function definitions: The term $\ilLetRecA{f}{\slist{x}}{s}{t}{X}$ is annotated with a set of variables $X$.

\subsection{Inductive Definition of the Liveness Judgment}

We define inductively the judgment \ndef{live}, which characterizes sound results of a liveness analysis.
\begin{center}
\begin{tabular}{lcrll}
 \multirow{3}{*}{$\live{\LC}{\lv}{s}$}&
 \multirow{3}{*}{~~where~~}
 &$\LC$&$:~\textit{context}\,(\M{set}\,\Var)$&~liveness for functions\\
 &&$\lv$&$:~\M{set}\,\Var$&~live variables\\
 &&$s$&$:~\Exp$&~expression
\end{tabular}
\end{center}
The predicate $\live{\LC}{\lv}{s}$ can be read as
\textit{$\lv$ contains all variables significant to $s$ in any context satisfying the assumptions $\LC$.}
The context~$\LC$ records for every function $f$ a set of variables $\lv$ that we call the \ndef{globals} of $f$.
Assuming $\slist{x}$ are the parameters of $f$, we will arrange things such that the set $\lv\cup\slist{x}$ contains all variables significant for the body of $f$,
but never a parameter of $f$: $X\cap\slist{x}=\emptyset$.
Throughout the paper, $\LC$ is always a (partial) mapping from labels to globals,
and $\lv$ denotes a set of variables.

\begin{figure}[ht]
\begin{center}
  \begin{topprooftree}{Live-Op}
    \AxiomC{$\freenam{\extexp}\incl\lv$}
    \noLine
    \UnaryInfC{$\lv'\setminus\set{x}\incl\lv$}
    \AxiomC{$x\in\lv'$}
    \noLine
    \UnaryInfC{$\live{\LC}{\lv'}{s}$}
    \BinaryInfC{$\live{\LC}{\lv}{\ilLet{x}{\extexp}{s}}$}
  \end{topprooftree}
  \begin{topprooftree}{Live-Exp}
    \AxiomC{$\freenam{e}\incl\lv$}
    \UnaryInfC{$\live{\LC}{\ann{\lv}}{\ilReturn{e}}$}
  \end{topprooftree}
  \begin{topprooftree}{Live-App}
    \AxiomC{$\lv_1\incl\lv$}
    \AxiomC{$\freenam{\slist{e}}\incl\lv$}
    \BinaryInfC{$\live{\LC;f:\lv_1;\LC'}{\lv}{\ilGoto{f}{\slist{e}}}$}
  \end{topprooftree}
\end{center}
\begin{center}
  \begin{topprooftree}{Live-Cond}
    \AxiomC{$\freenam{e}\incl\lv$}
    \noLine
    \UnaryInfC{$\lv_1\cup\lv_2\incl\lv$}
    \AxiomC{$\live{\LC}{\lv_1}{s_1}$}
    \noLine
    \UnaryInfC{$\live{\LC}{\lv_2}{s_2}$}
    \BinaryInfC{$\live{\LC}{\lv}{\ilIf{e}{s_1}{s_2}}$}
  \end{topprooftree}
  \begin{topprooftree}{Live-Fun}
    \Axiom$\fCenter\live{\LC;f:\lv_1}{\lv_1\cup\slist{x}}{s_1}$
    \noLine
    \UnaryInf$\fCenter\live{\LC;f:\lv_1}{\lv_2}{s_2}$
    \Axiom$\fCenter\lv_1\cap\slist{x}=\emptyset$
    \noLine
    \UnaryInf$\fCenter\lv_2\incl\lv$
    \BinaryInfC{$\live{\LC}{\lv}{\ilLetRecA{f}{\slist{x}}{s_1}{s_2}{X_1}}$}
  \end{topprooftree}
\end{center}
\caption{Liveness: An approximation of the significant variables for IL/I}
\label{fig:liveness}
\end{figure}
\subsubsection{Description of the Rules.}
\nrule{Live-Op}
 ensures that all variables free in $\eta$ are live.
Every live variable of the continuation $s$ except $x$ must be live at the assignment.
We require $x$ to be live in the continuation.
\nrule{Live-Cond} ensures that the live variables of a conditional at least contain the free variables of the condition, and the variables live in the consequence and alternative.
\nrule{Live-Exp} ensures that for programs consisting of a single expression~$e$ at least the free variables of~$e$ are live.
\nrule{Live-App} ensures that the free variables of every argument are live, and that the globals $\lv_1$ of~$f$ are live at the call site.
\nrule{Live-Fun} records the annotation $X_1$ as globals for $f$ in $\LC$, ensures that $X_1\cup\slist{x}$ is a large enough live set for the function body, and that $X_1$ does not contain parameters of~$f$.
The live variables~$X_2$ of the continuation~$t$ must be live at the function definition.

\begin{theorem}[Liveness is Decidable]
  For all $\LC$, $X$ and annotated $s$, it is efficiently decidable whether $\live{\LC}{X}{s}$ holds.
\label{thm:live_dec}
\end{theorem}
The proof of \myref{thm:live_dec} is constructive and yields an efficient, extractable decision procedure.
The decision procedure recursively descends on the program structure,
checking the conditions of the appropriate rule in every step.

\subsection{Liveness Approximates Significance}
We show that the live variables approximate the significant variables.
We write $\liveCtx{L}{\LC}$ if a context $L$ satisfies the assumptions~$\LC$,
and define:

\begin{center}
  \begin{topprooftree}{LiveCtx1}
    \AxiomC{$\liveCtx{L}{\LC}$}
    \AxiomC{$\lv\cap\slist{x}=\emptyset$}
    \AxiomC{$\live{\LC;f: \lv}{\lv\cup\slist{x}}{s}$}
      \TrinaryInfC{$\liveCtx{L;f:(\slist{x},s)}{\LC;f:\lv}$}
  \end{topprooftree}
  \begin{topprooftree}{LiveCtx2}
    \AxiomC{$$}
    \UnaryInfC{$\liveCtx{\emptyset}{\emptyset}$\phantom{$f()$}}
  \end{topprooftree}
\end{center}
\nbrule{LiveCtx1} ensures that $X$ does not contain parameters and that $X\cup\slist{x}$ is a large enough live set for the function body $s$ under the context $\LC;f: \lv$.

We can now formally state the soundness of the live predicate.
We prove that if $\live{\LC}{\lv}{s}$, then~$\lv$ contains at least the significant variables of $s$ in every context $L$ that satisfies the assumptions~$\LC$. We write $V=_{\lv}V'$ if $V$ and $V'$ agree on $X$, that is if $\forall x\in{X}, V x = V' x$.

\begin{theorem}
For every program $s$, if $\lives{\LC}{\lv}{s}$ and $\liveCtx{L}{\LC}$ and $V=_{\lv}V'$, then
{$(L,V,s)_I \PrefixEq (L,V',s)_I$}.
\label{thm:live_I}\end{theorem}

\section{Coherence}
\label{chap:invariance}
\label{sec:coherence}
Coherence is a syntactic condition that ensures that a program is invariant.
Coherence is defined relative to liveness information $\lives{\LC}{\lv}{s}$.

In the following programs, the set of globals of $f$ is \set{x}.
The program on the left is not invariant, while the program on the right is coherent.
\begin{center}
\begin{minipage}{.45\textwidth}
\begin{lstlisting}[label=lst:diff_sem,caption={}]
let x = 7 in
fun f () : {x} = x in
let x = 5 in f ()
\end{lstlisting}
\end{minipage}
\begin{minipage}{.45\textwidth}
\begin{lstlisting}[label=lst:invariant,caption={}]
let x = 7 in
fun f () : {x} = x in
let y = 5 in f ()
\end{lstlisting}
\end{minipage}
\end{center}
In the program on the left in line~3, the value of $x$ is $5$ and disagrees with the value of $x$ in the closure of $f$.
In the program on the right, $x$ was not redefined, hence both IL and IL/I will compute 7.
We say a function $f$ is \emph{available} as long as none of $f$'s globals were redefined.
The inductive definition of coherence ensures only available functions are applied.

\subsection{Inductive Predicate}
The coherence judgment is of the form \fbox{$\coh{\LT}{a}{s}$}, where
$s$ is an annotated program and
$\LT$ is similar to the context in the liveness judgment.
We exploit that contexts realize a partial mapping,
and maintain the invariant that $\LT$
maps only \emph{available} functions to their globals, and all other functions to $\bot$.
The inductive definition given below ensures that only available
functions are applied.

\begin{center}
\small
  \begin{topprooftree}{Coh-Op}
    \AxiomC{$\coh{\restrict{\LT}{\Var\setminus\set{x}}}{a}{s}$}
    \UnaryInfC{$\coh{\LT}{\anni{\lv}{a}}{\ilLet{x}{\eta}{s}}$}
  \end{topprooftree}
  \begin{topprooftree}{Coh-Exp}
    \AxiomC{$$}
    \UnaryInfC{$\coh{\LT}{\ann{\lv}}{\ilReturn{e}}$}
  \end{topprooftree}
  \begin{topprooftree}{Coh-App}
    \AxiomC{$\LT f \not= \bot$}
    \UnaryInfC{$\coh{\LT}{\ann{\lv}}{\ilGoto{f}{\slist{y}}}$}
  \end{topprooftree}
\end{center}
\begin{center}
\small
  \begin{topprooftree}{Coh-Cond}
    \AxiomC{$\coh{\LT}{a}{s}$}
    \AxiomC{$\coh{\LT}{b}{t}$}
    \BinaryInfC{$\coh{\LT}{\annii{\lv}{a}{b}}{\ilIf{x}{s}{t}}$}
  \end{topprooftree}
  \begin{topprooftree}{Coh-Fun}
    \AxiomC{$\coh{\LT;f:\lv}{b}{t}$}
    \AxiomC{$\coh{\restrict{\LT;f:\lv}{\lv}}{a}{s}$}
    \BinaryInfC{$\coh{\LT}{\annii{\lv}{a}{b}}{\ilLetRecA{f}{\slist{x}}{s}{t}{\lv}}$}
  \end{topprooftree}
\end{center}

\subsubsection{Description of the Rules.}
\nrule{Coh-Op} deals with binding a variable $x$.
Every function that has $x$ as a global (i.e. $x\in\LT f$) becomes unavailable, and must be removed from $\LT$.
We write $\restrict{\LT}{\lv}$ to remove all definitions from $\LT$ that require more globals than $\lv$. Trivally, $\restrict{\LT}{\Var}=\LT$.
To remove all definitions from $\LT$ that use $x$ as global, we use $\restrict{\LT}{\Var\setminus\set{x}}$.

Formally, the definition of \restrict{\LT}{X} exploits the list structure of contexts:
\begin{center}
\begin{minipage}{.4\textwidth}
\begin{align*}
 \restrict{\emptyset}{\lv}~&=~\emptyset\\
 \restrict{\LT;f:\bot}{\lv}~&=~\restrict{\LT}{\lv};f:\bot
\end{align*}
\end{minipage}
\begin{minipage}{.55\textwidth}
\begin{align*}
 \restrict{\LT;f:\lv'}{\lv}~&=~\restrict{\LT}{\lv};f:\lv' &&\lv'\incl\lv\\
 \restrict{\LT;f:\lv'}{\lv}~&=~\restrict{\LT}{\lv};f:\bot &&\lv'\not\incl\lv
\end{align*}
\end{minipage}
\end{center}
\nrule{Coh-App} ensures only available functions can be applied, since $\LT$ maps functions that are not available to~$\bot$.
\nrule{Coh-Fun} deals with function definitions.
When the definition of a function $f$ is encountered, its globals $X$  according to the annotation are recorded in~$\LT$.
In the function body~$s$, only functions that require at most $X$ as globals are available, so the
context is restricted to $\restrict{\LT;f:\lv}{\lv}$.

\begin{theorem}[Coherence is Decidable]
  For all $\LT$ and annotated~$s$, it is efficiently decidable whether $\coh{\LT}{a}{s}$ holds.
\label{thm:coh_dec}
\end{theorem}

\subsection{Coherent Programs are Invariant}
Given a configuration $(F, V,t)$ such that $F f = (V', \slist{x}, s)$, the \textbf{agreement invariant} describes a correspondence between the values of variables in the function closure $V'$ and the environment $V$.
If the closure of~$f$ is available, the closure environment~$V'$ agrees with the primary environment~$V$ on~$f$'s globals~$X$: {${V' =_{\lv} V}$}.
We write $\agr{F}{V}{\LT}$ if $\forall f\in\dom{F}\cap\dom{\LT},~ V' =_{\lv} V$ (where $\LT f = \lv$ and  $F f = (V',\slist{x}, s)$).

Function application continues evaluation with the function body from the closure.
Assume $F f = (V', \slist{x}, s)$ and consider the IL reduction:
$$    (F, V, \ilGoto{f}{\slist{e}}) \fevalsg{}{}
             (F^f,V'\update{\slist{x}}{\slist{v}}a, s)$$
If coherence is to be preserved, $s$ must be coherent under suitable assumptions.
We say $\LT$~approximates $\LT'$ if whenever $\LT f$ is defined, it agrees with~$\LT'$ and define
$\LT\preceq{}\LT'~:\!\iff~\forall{f}\in\dom{\LT},~\LT f = \LT' f$.
The \ndef{context coherence} predicate \cohCtx{F}{\LT} ensures that all function bodies in closures are coherent.
It is defined inductively on the context:
\begin{center}
  \begin{topprooftree}{CohC-Emp}
    \AxiomC{$\phantom{X}$}
    \UnaryInfC{$\cohCtx{\emptyset}{\emptyset}$}
  \end{topprooftree}
  \begin{topprooftree}{CohC-Bot}
    \AxiomC{$\cohCtx{F}{\LT}$}
    \UnaryInfC{$\cohCtx{F;f\!:\!b}{\LT;f\!:\!\bot}$}
  \end{topprooftree}
  \begin{topprooftree}{CohC-Con}
    \AxiomC{$\lives{\LC'}{\lv\cup\slist{x}}{s}~~\LT;f\!:\!\lv \preceq \LC' $}
    \noLine
    \UnaryInfC{$\coh{\restrict{\LT;f:\lv}{\lv}}{a}{s}~ \cohCtx{F}{\LT}$}
    \UnaryInfC{$\cohCtx{F;f:(V,\slist{x},s)}{\LT,f:\lv}$}
  \end{topprooftree}
\end{center}
\nrule{CohC-Con} encodes two requirements:
First, the body of~$f$ must be coherent under the context restricted to the globals $\lv$ of $f$ (cf. \nrule{Coh-Fun}).
Second, $X\cup\slist{x}$ must suffice as live variables for the function body $s$ under some assumptions~$\LC'$ such that $\LT;f:\lv$ approximates $\LC'$.
Approximation ensures stability under restriction: $\cohCtx{F}{\LT}\Rightarrow\cohCtx{F}{\restrict{\LT}{X}}$.

We define $\M{strip}(V,\slist{x},s)~=~(\slist{x},s)$ and lift \M{strip} pointwise to contexts.

\begin{theorem}[Coherence implies Invariance]
Let $\coh{\LT}{\lv}{s}$ and $\cohCtx{F}{\LT}$ and $\lives{\LC'}{\lv}{s}$ such that $\LT\preceq\LC'$.
Then for all $V=_{\lv}V'$ such that $\agr{F}{V}{\LT}$, it holds
$(F,V,s)_F\PrefixEq(\M{strip}\,F, V', s)_I$.
\label{cor:main_thm}
\end{theorem}
\myref{cor:main_thm} reduces the problem of translating between IL/I and IL to the problem of establishing coherence.
For the translation from IL to IL/I, it suffices to establish coherence while preserving IL semantics.
Since SSA and functional programming correspond~\cite{Kelsey:1995, DBLP:journals/sigplan/Appel98},
the translation from IL/I to IL can be seen as~SSA construction~\cite{DBLP:journals/toplas/CytronFRWZ91}, and the translation from IL to IL/I, which we treat in the next section, as SSA destruction.

\section{Translating from IL/F to IL/I via Coherence}
\label{trans:fct}
\label{sec:destr}
\label{sec:rassign}
\label{sec:il:funcoh}
 The simplest method to establish coherence while preserving IL semantics is $\alpha$-renaming the program apart.
A renamed-apart program (for formal definition see \myref{sec:renamedapart}) is coherent, since every function is always available.
The properties of $\alpha$-conversion ensure semantic equivalence.

We present an algorithm that establishes coherence and uses no more different names than the maximal number of simultaneously live variables in the program.
This algorithm corresponds to the assignment phase of SSA-based register allocation~\cite{DBLP:conf/cc/HackGG06}. The algorithm requires a renamed-apart program as input to ensure that every consistent renaming can be expressed as a function from $\Var\to\Var$.
We proceed in two steps:
\begin{enumerate}
\item We define the notion of \emph{local injectivity} for a function $\rho:\Var\to\Var$.
      We show that renaming with a locally injective $\rho$ yields an $\alpha$-equivalent and coherent program $\rho\,s$.
\item We give an algorithm $\mathit{rassign}$ and show that it constructs a locally injective $\rho$ that uses the minimal number of different names.
\end{enumerate}

\newcommand{\sterm}{\ensuremath{s}}
\newcommand{\tterm}{\ensuremath{t}}
We introduce \textbf{more liveness annotations} before every term in the syntax, i.e. wherever a term $s$ appeared before, now a term $\inlan{X}s$ appears that annotates $s$ with the set $X$.
From now on, $\sterm,\tterm$ range over such annotated terms.
We define the projection $[\inlan{X}s]=X$.
The annotation corresponds directly to the live set parameter $\lv$ of the relation $\lives{\LC}{\lv}{s}$, hence it suffices to write $\livesA{\LC}{\lv}{s}$ for annotated programs.

\subsection{Local Injectivity}
We define inductively a judgment $\ri{\rho}{a}{\sterm}$ where $\rho:\Var\to\Var$ and $s$ is an annotated program.
We use the following notation for injectivity on~$X$:
$$\injon{f}{\lv}~:\!\iff~\forall{x}\,{y}\in\lv,~f\,x=f\,y\impl{}x=y$$
The rules defining the judgement are given below and require $\rho$ to be injective on every live set $X$ annotating any subterm:

\begin{center}
\small
  \begin{topprooftree}{Inj-Op}
    \AxiomC{$\injon{\rho}{\lv}$}
    \AxiomC{$\ri{\rho}{a}{s}$}
    \BinaryInfC{$\ri{\rho}{\anni{\lv}{a}}{\ilLetAA{x}{\eta}{s}{\lv}}$}
  \end{topprooftree}
  \begin{topprooftree}{Inj-Val}
    \AxiomC{$\injon{\rho}{\lv}$}
    \UnaryInfC{$\ri{\rho}{\ann{\lv}}{\ilReturnAA{e}{\lv}}$}
  \end{topprooftree}
  \begin{topprooftree}{Inj-App}
    \AxiomC{$\injon{\rho}{\lv}$}
    \UnaryInfC{$\ri{\rho}{\ann{\lv}}{\ilGotoAA{f}{\slist{y}}{\lv}}$}
  \end{topprooftree}
\end{center}
\begin{center}
\small
  \begin{topprooftree}{Inj-Cond}
    \AxiomC{$\injon{\rho}{\lv}$}
    \AxiomC{$\ri{\rho}{a}{\sterm}$}
    \AxiomC{$\ri{\rho}{b}{\tterm}$}
    \TrinaryInfC{$\ri{\rho}{\annii{\lv}{a}{b}}{\ilIfAA{x}{\sterm}{\tterm}{\lv}}$}
  \end{topprooftree}
  \begin{topprooftree}{Inj-Fun}
    \AxiomC{$\injon{\rho}{\lv}$}
    \AxiomC{$\ri{\rho}{a}{s}$}
    \AxiomC{$\ri{\rho}{b}{\tterm}$}
    \TrinaryInfC{$\ri{\rho}{\annii{\lv}{a}{b}}{\ilLetRecAA{f}{\slist{x}}{s}{\tterm}{\lv}{X_1}}$}
  \end{topprooftree}
\end{center}

Let $\Var_B(s)$ be the set of variables that occur in a binding position in~$s$, and \freenam{s} be the set of free variables of $s$.
For our theorems, several properties are required:
\begin{enumerate}[label=(\arabic*)]
\item The program must be without unreachable code, i.e.
in every subterm $\ilLetRec{f}{\slist{x}}{s}{t}$ it must be the case that $f$ is applied in $t$.
\item A variable in $\Var_B(s)$ must not occur in a set of globals in $\LT$.
We define $\LT\incl{U}~:\!\iff~\forall {f}\in\dom\LT,~\LT\,f\incl{U}$.
\item A variable in $\Var_B(s)$ must not occur in the annotation $[s]$.
We write $s\incl{U}$ if for every subterm $t$ of $s$ it holds that every $x\in[t]$ is either
in $U$ or bound at $t$ in $s$.
\end{enumerate}
For renamed-apart programs, these conditions ensure that the live set $X$ in \nrule{Inj-Fun} always contains the globals $X_1$ of $f$ (cf. \nrule{Live-App}).

\begin{theorem}
\label{thm:inj_coh}
Let $s$ be a renamed-apart program without unreachable code such that $\livesA{\LC}{\lv}{s}$,
$\LC\incl\freenam{s}$ and $s\incl\freenam{s}$. Then
\begin{align*}
\ri{\rho}{a}{s}
~\impl~\coh{\rho~(\restrict{\LC}{[s]})}{\rho\,a}{(\rho\,s)}
\end{align*}
\end{theorem}
\myref{thm:inj_coh} states that the renamed program $\rho\,s$ is coherent under the assumptions $\rho~(\restrict{\LC}{[s]})$, i.e. the point-wise image of $\restrict{\LC}{[s]}$ under $\rho$.

Renaming with a locally injective renaming produces an $\alpha$-equivalent program (for formal definition see \myref{sec:alphaequiv}), and hence preserves program equivalence:

\begin{theorem}
\label{thm:inj_alpha}
Let $\sterm$ be a renamed-apart program without unreachable code such that $\livesA{\LC}{a}{\sterm}$, $\LC\incl\freenam{s}$ and $s\incl\freenam{s}$.
Let $\rmp,\rmpp:\Var\to\Var$ such that $\rmp$ is the inverse of $\rmpp$ on $\freenam{s}$. Then
$\ri{\rho}{a}{\sterm}~\impl~\alphai{\rmp}{\rmpp}{\rho\,\sterm}{\sterm}$
\end{theorem}

\newcommand{\ra}[3]{\ensuremath{\mathit{rassign}\,#3\,#1}}
\newcommand{\freshv}{\ensuremath{\mathit{fresh}}}
\newcommand{\fresh}[1]{\ensuremath{\freshv\,#1}}
\newcommand{\freshl}[2]{\ensuremath{\mathit{freshlist}\,#1\,#2}}

\subsection{A Simple Register Assignment Algorithm}
\label{sec:ra_algo}

The algorithm \textup{rassign} is parametrized by a function $\freshv:\M{set}\,\Var\to\Var$ of which we require $\freshv{\lv}\not\in\lv$ for all finite sets of variables $\lv$.
Based on $\freshv$, we define a function $\freshl{\lv}{n}$ that yields a list of $n$ pairwise-distinct variables such that $(\freshl{\lv}{n})\cap\lv=\emptyset$.
The SSA algorithm must process the program in an order compatible with the dominance order to work~\cite{DBLP:conf/cc/HackGG06}.
In our case it suffices to simply recurse on $s$ as follows:
\[
\begin{array}{lrl}
\ra{(\ilLetAA{x}{\eta}{s}{X})}{(\anni{\G}{a})}{\rho}&=&\ra{s}{a}{(\rho[x\mapsto y])}\\
\multicolumn{3}{l}{\quad~\textup{where}~y=\fresh{(\rho ([s]\setminus\set{x}))}}\\
\ra{(\ilIfAA{e}{\sterm}{\tterm}{X})}{(\annii{\G}{a}{b})}{\rho}&=&\ra{t}{b}{(\ra{s}{a}{\rho})}\\%
\ra{(\ilReturnAA{e}{X})}{\ann{\G}}{\rho}&=&\rho\\
\ra{(\ilGotoAA{f}{\slist{e}}{X})}{\ann{\G}}{\rho}&=&\rho\\
\ra{(\ilLetRecAA{f}{\slist{x}}{s}{\tterm}{X}{X'})}{(\annii{\G}{a}{b})}{\rho}&=&\ra{\tterm}{b}{(\ra{s}{a}{(\rho[\slist{x}\mapsto\slist{y}])})}\\
\multicolumn{3}{l}{\quad~\textup{where}~\slist{y}=\freshl{(\rho ([s]\setminus\slist{x}))}{|\slist{x}|}}
\end{array}
\]

We prove in \myref{thm:rassign_corr} that the algorithm is correct for any choice of $\mathit{fresh}$ and $\mathit{freshlist}$,
as long as they satisfy the specifications above.

\begin{theorem}
Let $s$ be renamed-apart such that $\livesA{\LC}{}{s}$, $\LC\incl\freenam{s}$ and $s\incl\freenam{s}$.
Let $\rho$ be injective on $[s]$.
Then: $ \ri{\ra{s}{a}{\rho}}{a}{s}$.

\label{thm:rassign_corr}
\end{theorem}

\newcommand{\occurvars}[1]{\ensuremath{\mathcal{V}_O(#1)}}
\newcommand{\segment}[1]{\ensuremath{\mathcal{S}(#1)}}

Our implementation of $\mathit{fresh}$ implements the heuristic of simply choosing the smallest unused variable.
\myref{thm:rassign_bound} shows that for this choice of $\mathit{fresh}$, the largest live set determines the number of required names.
We use $\segment{k}$ to denote the set of the $k$ smallest variables, and
$\occurvars{s}$ to denote the set of variables occurring (free or in a binding position) in $s$.
\begin{theorem}
Assume $\mathit{fresh}\,X$ yields a variable less or equal to $|X|$.
Let $\sterm$ be renamed-apart such that $\livesA{\LC}{X}{\sterm}$, $\LC\incl\freenam{s}$ and $s\incl\freenam{s}$.
Let $k$ be the size of the largest set of live variables in $\sterm$, and $\ra{\sterm}{a}{\rho} = \rho'$.
If $\rho (\freenam{s}) \incl \segment{n}$ then
$\rho' (\occurvars{s}) \incl \segment{\mathit{max}\set{n,k}}$.
\label{thm:rassign_bound}
\end{theorem}
We prove a slightly generalized version of \myref{thm:rassign_bound} by induction on $s$.

\section{Formal Coq Development}
\label{sec:formdev}
Each theorem and lemma in this paper is proven as part of a larger Coq development,
which is available online\footnote{\url{http://www.ps.uni-saarland.de/~sdschn/publications/lvc15}}.
The development extracts to a simple compiler that, for instance, produces program~(b) when given program~(a) from the introduction as input.

The formalization uses De-Bruijn representation for labels, and named representation for variables.
Notable differences to the paper presentation concern the treatment of annotations, the technical realization of the definition of liveness, and the inductive generalizations of Theorems~6-9.

 \section{Conclusion}
We presented the functional intermediate language IL and developed the notion of coherence, which
provides for a canonical and verified translation between functional and imperative programs.
We formulated an register assignment algorithm by recursion on the structure of IL that achieves the same bound on the number of required registers as SSA-based register assignment.
Coherence allowed us to justify correctness without directly arguing about program semantics by proving that the algorithm $\alpha$-renames to a coherent program.

\enlargethispage{6mm}

\label{chap:conclusion}
\label{sec:conclusion}

\section{Appendix}

\subsection{Table of Variable Names and Types}

\begin{tabular}{lll}
\textbf{Variable}&\textbf{Type}&\textbf{comment}\\
$\Val$ & set & set of values\\
$\beta$ & $\Val\to\set{0,1}$ & conversion to truth value\\
$v$ & $\Val$ & value\\
$\Exp$ & set & set of expressions\\
$\Var$ & set & set of variables\\
$e$ & $\Exp$ & expression\\
$x,y,z$ & $\Var$ & variables\\
$\Fun$ & set & set of lables\\
$f,g$ & $\Fun$ & labels\\
$\Act$ & set & set of actions  \\
$\eta$ & $\Exp + \Act$  & extended expression \\
$\alpha$ & $\Act$ & action  \\
$\Trm$ & set & set of terms\\
$s,t$ & $\Trm$ & terms\\
$V$ & $\Var\to\Valb$ & environment\\
$\closures$ & set & set of closures\\
$F$ & context of $\closures$ &\\
$\Evt$ & set & set of events\\
$\phi$ & $\Evt$ & event\\
$\tau$ & $\Evt$ & silent event\\
$\mathcal{B}$ & set & set of blocks \\
$L$ & context of $\mathcal{B}$ & \\
$\Sigma$ & set & set of states (LTS)\\
$\sigma$ & $\Sigma$ & state, configuration\\
$\Pi$ & set & set of partial traces\\
$\pi$ & $\Pi$ & partial trace \\
$\nil$ & $\Pi$ & empty trace\\

\end{tabular}

\subsection{$\alpha$-Equivalence}
\label{sec:alphaequiv}
We formalize a generalization of alpha equivalence as an inductively defined judgment
$\alphai{\rmp}{\rmpp}{s}{t}$ where $\rmp,\rmpp:\Var\to\Var$ and $s,t$ are terms.
The mapping~$\rmp$ describes how the free variables of $s$ map to free variables of~$t$,
and~$\rmpp$ describes how the free variables of~$t$ map to free variables of~$s$.
If $\alphai{\rmp}{\rmpp}{s}{t}$ holds, then $\rmpp$ is the inverse of $\rmp$ on
$\freenam{s}$, i.e. $$\forall x\in\freenam{s},\,\rmpp(\rmp\,x) = x$$
Symmetrically, $\rmp$ is the inverse of $\rmpp$ on
$\freenam{t}$.

The formalization assumes a similar judgment $\alphaie{\rmp}{\rmpp}{e}{e'}$ for $\alpha$-equivalence of expressions.
The variable case of judgment for expressions explains how $\rmp$ and $\rmpp$ are used:
\begin{center}
\begin{topprooftree}{Alpha-Var}
  \AxiomC{$\rmp x = y$}
  \AxiomC{$\rmpp y = x$}
  \BinaryInfC{$\alphaie{\rmp}{\rmpp}{x}{y}$}
\end{topprooftree}
\end{center}
\nrule{Alpha-Var} ensures that $\rmp$ maps $x$ to $y$ and $\rmpp$ maps $y$ to $x$.

The other rules of the expression judgment are structurally recursive and we omit them.

\begin{figure}[htb]
\begin{center}
  \begin{topprooftree}{Alpha-Op}
    \AxiomC{$\alphaie{\rmp}{\rmpp}{\eta}{\eta'}$}
    \AxiomC{$\alphai{\rmp[x\mapsto x']}{\rmpp[x'\mapsto x]}{s}{s'}$}
    \BinaryInfC{$\alphai{\rmp}{\rmpp}{\ilLet{x}{\eta}{s}}{\ilLet{x'}{\eta'}{s'}}$}
  \end{topprooftree}
  \begin{topprooftree}{Alpha-Val}
    \AxiomC{$\alphaie{\rmp}{\rmpp}{e}{e'}$}
    \UnaryInfC{$\alphai{\rmp}{\rmpp}{\ilReturn{e}}{\ilReturn{e}}$}
  \end{topprooftree}
\end{center}
\begin{center}
  \begin{topprooftree}{Alpha-App}
    \AxiomC{$\forall i,~ \alphaie{\rmp}{\rmpp}{e_i}{e'_i}$}
    \UnaryInfC{$\alphai{\rmp}{\rmpp}{\ilGoto{f}{\slist{e}}}{\ilGoto{f}{\slist{e'}}}$}
  \end{topprooftree}
  \begin{topprooftree}{Alpha-Cond}
    \AxiomC{$\alphaie{\rmp}{\rmpp}{e}{e'}$}
    \AxiomC{$\alphai{\rmp}{\rmpp}{s}{s'}$}
    \noLine
    \UnaryInfC{$\alphai{\rmp}{\rmpp}{t}{t'}$}
    \BinaryInfC{$\alphai{\rmp}{\rmpp}{\ilIf{e}{s}{t}}{{\ilIf{e'}{s'}{t'}}}$}
  \end{topprooftree}
\end{center}

\begin{center}
  \begin{topprooftree}{Alpha-Fun}
    \AxiomC{$\alphai{\rmp[\slist{x}\mapsto\slist{x'}]}{\rmpp[\slist{x'}\mapsto\slist{x}]}{s}{s'}$}
    \AxiomC{$\alphai{\rmp}{\rmpp}{t}{t'}$}
    \AxiomC{$|\slist{x}| = |\slist{x'}|$}
    \TrinaryInfC{$\alphai{\rmp}{\rmpp}{\ilLetRec{f}{\slist{x}}{s}{t}}{\ilLetRec{f}{\slist{x'}}{s'}{t'}}$}
  \end{topprooftree}
\end{center}

\caption{Inductive judgment generalizing $\alpha$-equivalence}
\label{fig:alpha}
\end{figure}

The relation has several pleasant properties.
\begin{lemma}[Reflexivity]
$\alphai{\mathit{id}}{\mathit{id}}{s}{s}$
\end{lemma}

\begin{lemma}[Symmetry]
$\alphai{\rmp}{\rmpp}{s}{s'}\Rightarrow \alphai{\rmpp}{\rmp}{s'}{s}$
\end{lemma}

\begin{lemma}[Transitivity]
$\alphai{\rmp_1}{\rmpp_1}{s}{s'}\Rightarrow \alphai{\rmp_2}{\rmpp_2}{s'}{s''} \Rightarrow
\alphai{\rmp_1\circ\rmp_2}{\rmpp_2\circ\rmpp_1}{s}{s'}
$
\end{lemma}

We validate our definition and prove soundness with respect to trace equivalence $\PrefixEq$.
We define $$V =_{\rmp,\rmpp} V' ~:\!\iff~ \forall x y, \rmp x = y \Rightarrow \rmpp y = x \Rightarrow V x = V' y$$
We relate two closures in the following way:
\begin{align*}
 &(V,\slist{x},s) =_\alpha (V',\slist{x'},s')
\\ ~:\!\iff~ &|\slist{x}| = |\slist{x'}| ~\land \exists \rmp\, \rmpp,~ V =_{\rmp,\rmpp} V' \land \alphai{\rmp[\slist{x}\mapsto\slist{x'}]}{\rmpp[\slist{x'}\mapsto\slist{x}]}{s}{s'}
\end{align*}
We then lift $=_\alpha$ point-wise to contexts of the same length.

\begin{theorem}
If $F =_\alpha F'$ and $V =_{\rmp,\rmpp} V'$
then $(F,V,s)\PrefixEq(F',V',s')$.
\end{theorem}

In the formal development we have an additional formalization of IL which uses De-Bruijn representation also for variables (and not just for labels).
We give a translation from the named IL to De-Bruijn IL, and prove this translation correct with respect to trace equivalence.
We then show that terms that are $\alpha$-equivalent by our inductive definition translate to identical terms in De-Bruijn representation.

 \newcommand{\apart}[3]{\ensuremath{#1\vdash#2\,\textbf{apart}\, #3}}

\newcommand{\vs}{\ensuremath{X}}

\subsection{Definition of Renamed Apart}
\label{sec:renamedapart}
A program is renamed apart, if every variable $x$ occurring in a binding position does not occur free and $x$ is different from every variable occurring in a different binding position.
We formulate an inductive predicate \apart{\vs}{s}{\vs'} that ensures this property.
The predicate maintains the invariant that all free variables of $s$ are in $\vs$,
and that $\vs'$ contains exactly the variables occurring in binding positions in $s$.

\begin{figure}[htb]
\begin{center}
  \begin{topprooftree}{Apart-Op}
    \AxiomC{$\freenam{e}\incl\vs$}
    \AxiomC{$\apart{\vs \cup \set{x}}{s}{\vs'}$}
    \BinaryInfC{$\apart{\vs}{\ilLet{x'}{\eta'}{s'}}{\vs' \cup \set{x}}$}
  \end{topprooftree}
  \begin{topprooftree}{Apart-Val}
    \AxiomC{$\freenam{e}\incl\vs$}
    \UnaryInfC{$\apart{\vs}{\ilReturn{e}}{\emptyset}$}
  \end{topprooftree}
\end{center}
\begin{center}
  \begin{topprooftree}{Apart-App}
    \AxiomC{$\freenam{\slist{e}}\incl \vs$}
    \UnaryInfC{$\apart{\vs}{\ilGoto{f}{\slist{e}}}{\emptyset}$}
  \end{topprooftree}
  \begin{topprooftree}{Apart-Cond}
    \AxiomC{$\freenam{e}\incl\vs$}
    \noLine
    \UnaryInfC{$\vs_s \cap \vs_t = \emptyset$}
    \AxiomC{$\apart{\vs}{s}{\vs_s}$}
    \noLine
    \UnaryInfC{$\apart{\vs}{t}{\vs_t}$}
    \BinaryInfC{$\apart{\vs}{\ilIf{e}{s}{t}}{\vs_s \cup \vs_t}$}
  \end{topprooftree}
\end{center}

\begin{center}
  \begin{topprooftree}{Apart-Fun}
    \AxiomC{$\apart{\vs}{t}{\vs_t}$}
    \noLine
    \UnaryInfC{$\apart{\vs \cup \slist{x}}{s}{\vs_s}$}
    \AxiomC{$\textup{unique}\,\slist{x}$}
    \AxiomC{$\slist{x} \cap \vs = \emptyset$}
    \noLine
    \UnaryInfC{$(\vs_s \cup \slist{x}) \cap \vs_t = \emptyset$}
    \TrinaryInfC{$\apart{\vs}{\ilLetRec{f}{\slist{x}}{s}{t}}{\vs_s\cup \vs_t \cup \slist{x}}$}
  \end{topprooftree}
\end{center}

\caption{Inductive definition of renamed apart}
\label{fig:apart}
\end{figure}

\begin{lemma}[Disjoint]
If $\apart{\vs}{s}{\vs'}$ then $\vs \cap \vs' = \emptyset$.
\end{lemma}

\begin{lemma}[Relation to free and bound variables]
If $\apart{\vs}{s}{\vs'}$ and then $\freenam{s}\incl\vs$ and $\vs' = \Var_B(s)$.
\end{lemma}

\subsection{A Procedure to Rename Apart}

\renewcommand{\ra}[3]{\mathit{apart}\,#1\,#2\,#3}

We define the procedure $$\mathit{apart}:(\Var\to\Var)\to(\M{set}\,\Var)\to\Exp\to(\M{set}\,\Var)\times\Exp$$ such that $\ra{\rho}{X}{s}=(X,s')$ ensures $s'$ is renamed apart and $\alpha$-equivalent to $s$. $X'$ contains the newly chosen variables now occurring in binding positions in $s'$.
\myref{thm:apart_correct} and \myref{thm:apart_alpha} make these claims precise.

\[
\begin{array}{lrl}
\ra{\rho}{X}{(\ilLet{x}{\eta}{s})}&=&(X' \cup \set{y}, \ilLet{y}{\rho\,\eta}{s'})\\
\multicolumn{3}{l}{\quad~\textup{where}~(X', s') = \ra{(\rho[x\mapsto y])}{(X \cup \set{y})}{s}}\\
\multicolumn{3}{l}{\quad~\textup{where}~y = \fresh{X}}\\
\ra{\rho}{X}{(\ilIf{e}{\sterm}{\tterm})}&=&(X_s\cup X_t, \ilIf{(\rho\,e)}{s'}{t'})\\
\multicolumn{3}{l}{\quad~\textup{where}~(X_s, s') = \ra{\rho}{X}{s}}\\
\multicolumn{3}{l}{\quad~\textup{where}~(X_t, t') = \ra{\rho}{(X \cup X_s)}{t}}\\
\ra{\rho}{X}{\ilReturn{e}}&=&(\emptyset, \ilReturn{\rho e})\\
\ra{\rho}{X}{(\ilGoto{f}{\slist{e}})}&=&(\emptyset, \ilGoto{f}{(\rho\,\slist{e})})\\
\ra{\rho}{X}{(\ilLetRec{f}{\slist{x}}{s}{t})}&=&(X_s\cup X_t \cup \slist{y}, \ilLetRec{f}{\slist{y}}{s'}{t'})\\
\multicolumn{3}{l}{\quad~\textup{where}~\slist{y} = \freshl{X}{|\slist{x}|}}\\
\multicolumn{3}{l}{\quad~\textup{where}~(X_s, s') = \ra{(\rho[\slist{x}\mapsto\slist{y}])}{(X\cup \slist{y})}{s}}\\
\multicolumn{3}{l}{\quad~\textup{where}~(X_t, t') = \ra{\rho}{(X \cup X_s \cup \slist{y})}{t}}\\
\end{array}
\]

\begin{theorem}[\textnormal{\textit{apart}} renames apart]
Let $s$ be a program such that $\rho (\freenam{s}) \incl X$ and $\ra{\rho}{X}{s}=(X', s')$.
Then: $\apart{X}{s'}{X'}$.
\label{thm:apart_correct}
\end{theorem}

\begin{theorem}[Renaming apart respects $\alpha$-conversion]
\label{thm:apart_alpha}
Let $s$ be a program such that $\rho (\freenam{s}) \incl X$ and $\ra{\rho}{X}{s}=(X', s')$ and
let $\rmpp$ be inverse to $\rho$ on $\freenam{s}$.
Then $\alphai{\rho}{\rmpp}{s}{s'}$.
\end{theorem}

 \subsection{Joining the Parts}
This section describes how the theorems proven in this paper fit together in a compiler.
Assume that the compiler uses IL as an intermediate language, and now wants to produce code for an IL program $s$. The compiler procedes as follows:

\begin{enumerate}
 \item Rename $s_1$ apart, obtaining an $\alpha$-equivalent program $s_2$ (\myref{thm:apart_alpha}).
 \item Run the algorithm $\mathit{rassign}$ on $s_2$ to obtain a register assignment $\rho$.
       Theorem \myref{thm:rassign_corr} ensures $\rho$ is locally injective.
 \item Rename $s_2$ accoding to $\rho$ and obtain $s_3$, which is $\alpha$-equivalent
   (\myref{thm:inj_alpha}) and coherent (\myref{thm:inj_coh}) because $\rho$ is locally injective.
 \item Theorem \myref{cor:main_thm} ensures that $s_3$ can be seen equivalently as an IL/I program,
       hence the functional program $s_1$ has been translated to an imperative program $s_3$.
\end{enumerate}

\printbibliography

\end{document}